\documentclass[10pt,a4paper,twocolumn]{article}

\usepackage[margin=1.8cm]{geometry}
\usepackage{amsmath, amssymb}
\usepackage{graphicx}
\usepackage{helvet}
\usepackage{multicol}
\usepackage{titlesec}
\usepackage{xcolor}

\usepackage{hyperref}  

\usepackage[backend=biber,style=ieee,maxnames=6,doi=true]{biblatex}
\usepackage{siunitx}  
\usepackage{subfig}

\usepackage{listings}
\lstdefinestyle{spice}{
  basicstyle=\ttfamily\footnotesize, 
  keywordstyle=\color{blue},         
  commentstyle=\color{green!50!black}, 
  numbers=left,                      
  numberstyle=\tiny\color{gray},
  stepnumber=1,
  breaklines=true,                   
  frame=single,                      
  columns=fullflexible,
  captionpos=b   
}
\titleformat{\section}{\large\bfseries\color{blue!60!black}}{\Roman{section}.\quad}{0em}{}
\titleformat{\subsection}  {\normalsize\bfseries\color{blue!50!black}}{\Alph{subsection}.\quad}{0em}{}

\renewenvironment{abstract}
{
    \small
    \noindent\textbf{\textit{Abstract —}}
}
{}

\begin{document}

\twocolumn[
\begin{center}
 
    {\LARGE\bfseries End-to-End Modeling of a Volatile TiO\textsubscript{2} Memristor for\par}   
    {\LARGE\bfseries Neuromorphic Circuit Simulation\par}
    \vspace{4pt}
    
    {\small Lukas Endres$^{1}$, Hannes Töpfer$^{1}$, Michaela Blum$^{2}$, Hauke Honig$^{2}$, Peter Schaaf$^{2}$\par}
    \vspace{1pt}

    {\footnotesize $^{1}$ Chair Advanced Electromagnetics, Institute of Information Technology, Technische Universität Ilmenau\par}
    \vspace{1pt}
    {\footnotesize $^{2}$ Chair Materials for Electrical Engineering and Electronics, Institute of Materials Science and Engineering; Institute of Micro and Nanotechnology, Technische Universität Ilmenau\par}
    
\end{center}

\vspace{0.2cm}
]


\begin{abstract} Memristors are promising devices for applications such as non-volatile memory, neuromorphic computing, logic circuits, and analog signal processing. The development of such systems requires accurate simulations based on models that reproduce the electrical behavior of real devices under both continuous and pulsed excitation.
This work presents the development of a simulation environment for a volatile TiO\textsubscript{2}-based memristor. Experimental measurement data are analyzed to verify the memristive behavior of the device and to identify a suitable model. The model parameters are then optimized to match the measured characteristics.
The resulting model is implemented in SPICE and validated by comparing simulation results with measurement data. The comparison shows a good agreement between simulation and experiment, demonstrating that the developed model is suitable for reproducing the electrical behavior of the investigated memristor and can be applied in circuit-level simulations, as demonstrated by a leaky integrate-and-fire neuron.
\end{abstract}
\section{Introduction}
The memristor was first experimentally realized in 2008 by researchers at HP Labs \cite{strukov2008missing}, based on the theoretical concept proposed by Chua in 1971 \cite{chua1971memristor}. Chua introduced the memristor as the missing fourth fundamental circuit element. Although this classification remains debated \cite{abraham2018case}, memristors have attracted significant interest due to their electrically controllable resistance switching behavior. \newline

Memristors can exhibit either non-volatile or volatile switching characteristics. Non-volatile devices retain their resistance state after the excitation is removed and are widely investigated for memory applications, whereas volatile devices gradually relax toward their equilibrium state. This volatile behavior is particularly attractive for neuromorphic circuits.

The development of memristor-based circuits relies heavily on simulation. Accurate simulations require models that reproduce the electrical behavior observed in experimental measurements. The first widely used model was the linear ion drift model introduced by Strukov et al. \cite{strukov2008missing}. However, real devices exhibit various non-ideal effects that cannot be captured by idealized models. As a result, numerous modelling approaches have been proposed to describe different memristive technologies and application requirements. A comprehensive overview of memristor models is provided in \cite{yakopcic2012spice}.

In this work, a fabricated volatile TiO\textsubscript{2} memristor is investigated using experimental measurement data. The objective is to identify a suitable memristor model, determine its parameters from measurement data, and implement the resulting model in a SPICE simulation environment. This fitted model is validated by comparing simulated and measured device characteristics, and its applicability is demonstrated in a leaky integrate-and-fire neuron circuit.

\section{Model Development Workflow}
The proposed workflow consists of five key stages:
\begin{enumerate}
    \item \textbf{Data Assessment:}  
    Evaluation of measurement data to ensure its suitability for modelling (e.g. the presence of pinched hysteresis and identifiable thresholds), as discussed in Chapter~\ref{chap: Measurement Data}.
    
    \item \textbf{Model Selection:}  
    Identification of a mathematical model that describes the essential behavior observed in the memristor measurement data, as detailed in Chapters~\ref{chap: Structure of Memristor Models} and \ref{chap: Selected Memristor Model}.

    \item \textbf{Parameter Optimization:}  
    Adjusting the model parameters to ensure a close match between the simulated and measured characteristics.  
    This step is based on experimental data and aims to minimise the error between the model and reality, as presented in Chapter~\ref{chap: Parameter Fitting and Optimization}.

    \item \textbf{SPICE Implementation:}  
     Integration of the optimized model into a reusable SPICE subcircuit, as described in Chapter~\ref{chap: SPICE Modelling}

    \item \textbf{Validation} Validation of the SPICE model by comparing simulation results with experimental measurements, followed by its demonstration in an application using a leaky integrate-and-fire neuron, as presented in Chapter~\ref{chap: Model Evaluation}.
\end{enumerate}

\section{Measurement Data} \label{chap: Measurement Data}
The model is based on measurements of a fabricated memristor with a lateral Pt/TiO\textsubscript{2}/Ag structure. These measurements were performed using a Keithley 2450 source measure unit in a two-terminal setup with a compliance current of \SI{1}{\micro\ampere} to prevent device damage. The obtained data form the basis for parameter extraction.

Two types of electrical measurement were used: an I–V measurement to characterise the resistive switching behaviour under continuous voltage excitation and a pulse measurement to analyse the dynamic response and relaxation when no external bias was applied.

\subsection{I–V Measurement}

The current–voltage (I–V) characteristic is the key method for identifying memristive behaviour. Its typical feature is the pinched hysteresis loop, where both branches intersect at the origin — indicating that the device retains a memory of its resistance state \cite{chua2019everything}. This definition is phenomenological and independent of the material system or switching mechanism.

The shape of the hysteresis loop provides insight into the switching dynamics: a right curvature indicates increasing resistance and left curvature indicates decreasing resistance, while nearly linear segments occur either below the switching threshold or when the device is in a saturated state, meaning that its resistance no longer changes with further voltage variation.

\begin{figure}[hb!]
\centering
    \subfloat[]{\includegraphics[width=0.82\linewidth]{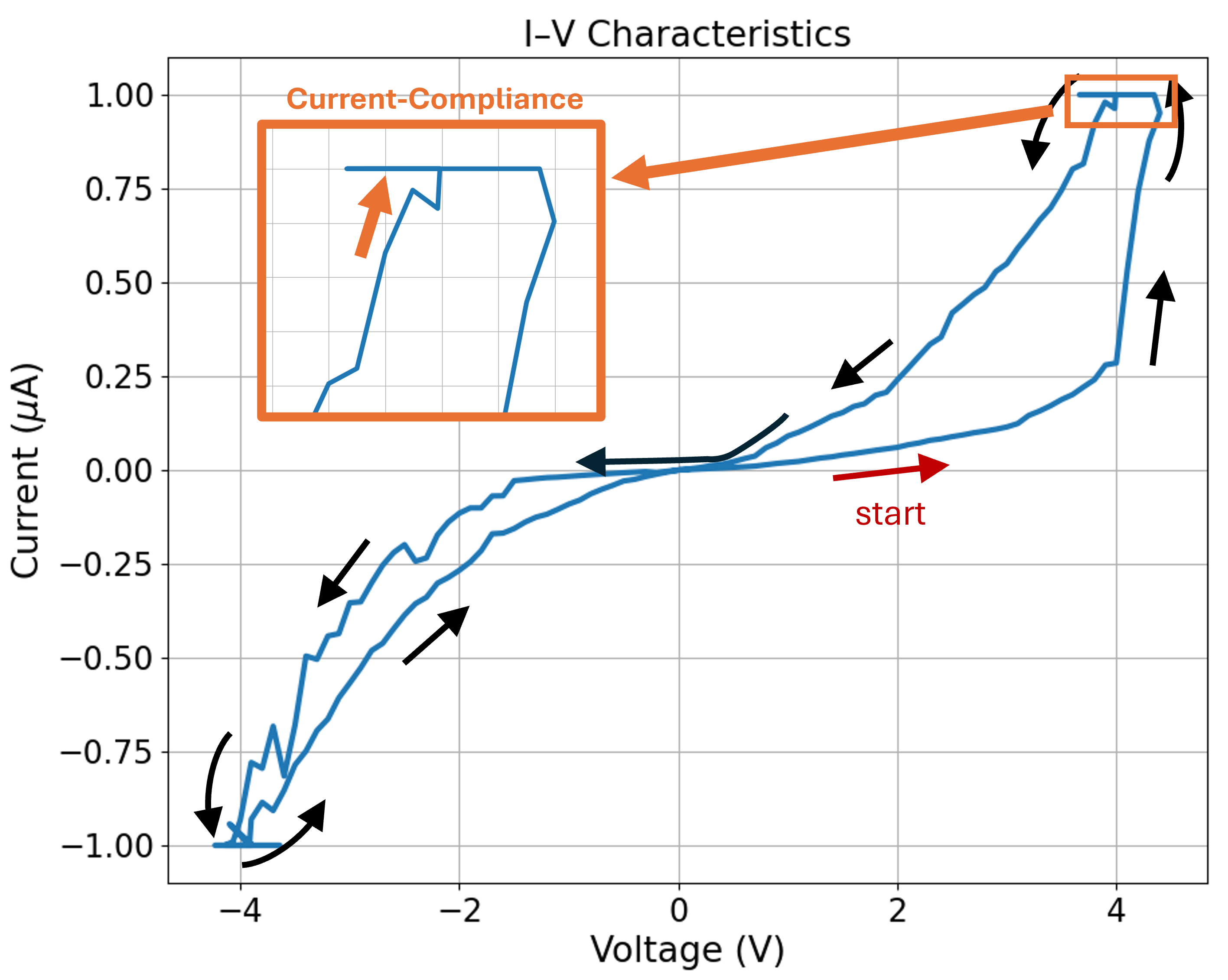}
    \label{pic: measurement_data_I_V}
    }\\[6pt]
    \subfloat[]{\includegraphics[width=0.82\linewidth]{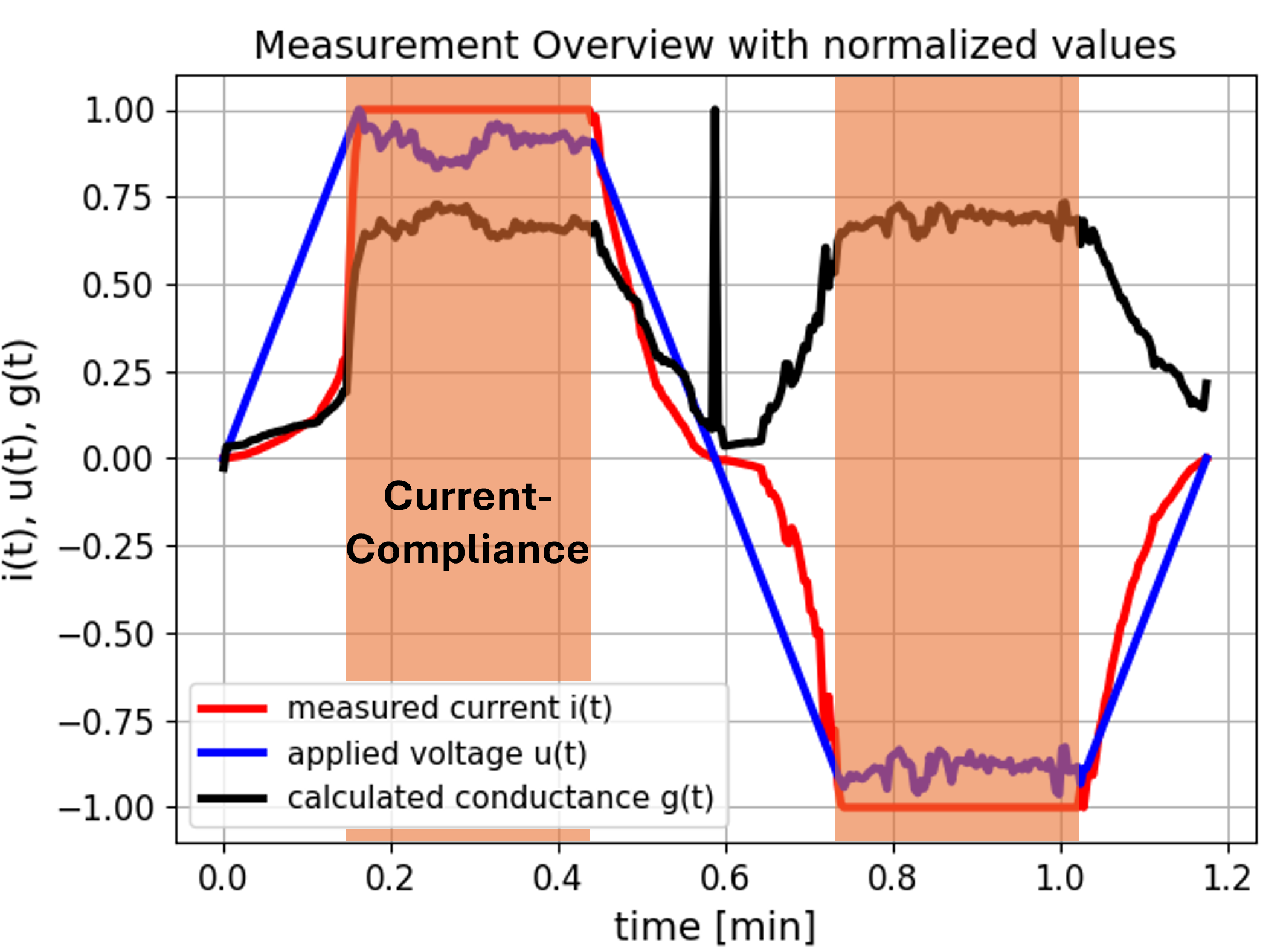}
    \label{pic: measurement_data_i_u_t}
    }
\caption{Measured data of the TiO\textsubscript{2}-based memristor: 
(a) I–V characteristic, (b) normalized representation of \(i(t)\), \(u(t)\), and \(g(t)\) during the I–V sweep.}
\label{fig:measurement_data_I_V_Messung}
\end{figure}

Figure~\ref{pic: measurement_data_I_V} shows the measured I–V characteristics of the device under a triangular voltage sweep from \SI{0}{\volt} to \SI{6}{\volt}, down to \SI{-6}{\volt}, and back to \SI{0}{\volt} (total measurement duration \SI{1.2}{\minute}), with arrows indicating the sweep direction. Figure~\ref{pic: measurement_data_i_u_t} presents the corresponding normalized signals \(u(t)\), \(i(t)\), and \(g(t)\).

\subsection{Compliance Current}

The compliance current (CC) defines the maximum current allowed through the memristor during measurement, primarily to prevent dielectric breakdown. In addition to providing protection, it also has a significant impact on switching.

The selected CC determines the lowest achievable resistance state \cite{zhang2016current}. Once the current limit is reached, filament growth halts, thereby fixing the device resistance. As reported in \cite{humood2021effect}, for filamentary devices the compliance current also influences the volatility of the conductive state:
low CC values (\SI{1}{\micro\ampere}–\SI{10}{\micro\ampere}) lead to volatile switching that relaxes within seconds, whereas higher values (\SI{500}{\micro\ampere}–\SI{1}{\milli\ampere}) result in stable, non-volatile states due to the formation of robust filaments.

When the current reaches the compliance limit, the measurement unit adjusts the voltage to keep the current constant (Figure~\ref{pic: measurement_data_i_u_t}). From this point onwards, the I–V curve is influenced by the instrument’s control loop rather than the device itself, leading to artifacts such as horizontal segments or fluctuating apparent resistance (Figure~\ref{pic: measurement_data_I_V}). These effects originate from the measurement system, not the intrinsic behavior of the memristor.

\subsection{Pulse Measurement}
\begin{figure}[b!]
\centering
    \subfloat[]{\includegraphics[width=0.97\linewidth]{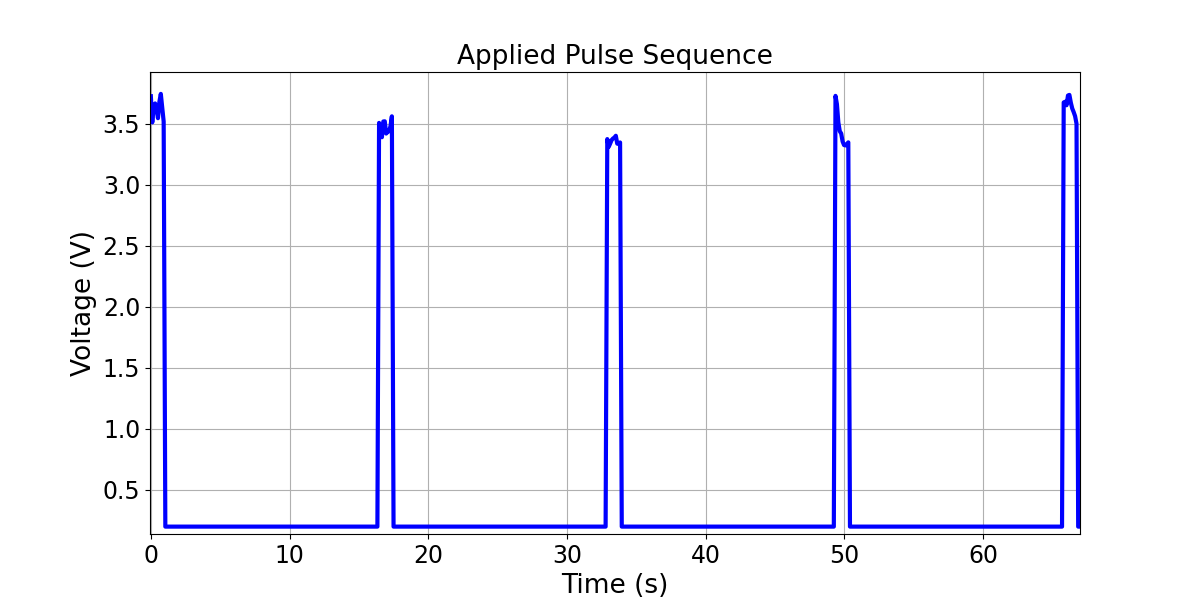}
    \label{pic: measurement_data_pulse_pulssequence}
    }\\[6pt]
    \subfloat[]{\includegraphics[width=0.97\linewidth]{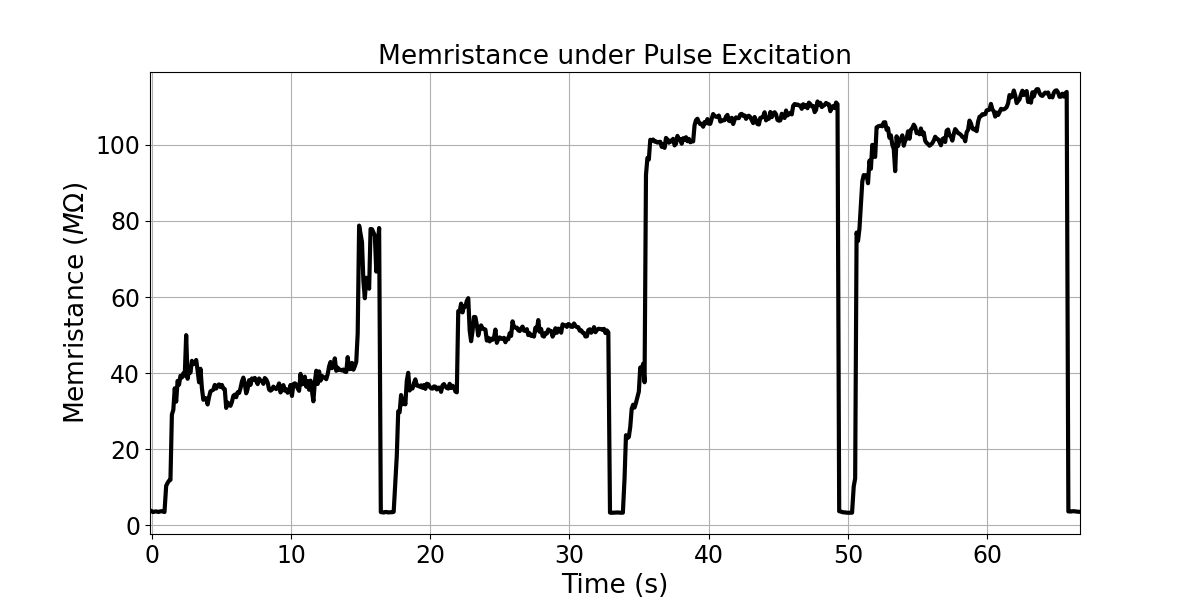}
    \label{pic: measurement_data_pulse_memristance}
    }
\caption{Measured resistive switching behavior under voltage pulse excitation: 
(a) applied pulse sequence, (b) memristance evolution over time.}
\label{fig:measurement_pulsmessung}
\end{figure}

In addition to the I–V characterization, pulse measurements were performed to study the resistive switching under repetitive voltage pulses.   
This approach enables the characterization of the volatile memristor's relaxation behavior. A positive SET voltage pulse exceeding the switching voltage identified from the I–V measurements increases the device resistance, whereas the READ voltage pulse remains below the switching threshold, allowing the resistance to relax back to its initial value while being monitored.\newline 
In contrast, non-volatile memristors retain their programmed resistance state after the applied voltage is removed. Subsequent switching therefore starts from this stored state, and the resistance can only be reduced by applying a negative voltage pulse.

For these measurements, a SET pulse of \SI{10}{\volt} with a duration of \SI{1.0}{\second} was applied, followed by a READ pulse of \SI{0.2}{\volt} lasting \SI{15}{\second}. This pulse sequence was repeated cyclically throughout the experiment. During the SET phase, the current reached the compliance current limit, causing the applied voltage to be capped between \SI{3.3}{\volt} and \SI{3.8}{\volt}. This behavior is illustrated in Figure~\ref{pic: measurement_data_pulse_pulssequence}.

\subsection{Evaluation of the Measurement Data}
The measured \textbf{I–V characteristics} exhibit the expected pinched hysteresis loop, confirming memristive behavior.  
Despite some noise, the resistance switches between \SI{4}{\mega\ohm} and \SI{80}{\mega\ohm}, yielding an \(R_\mathrm{max}/R_\mathrm{min}\) ratio of approximately 20.  
This limitation is mainly caused by the compliance current of \SI{1}{\micro\ampere}.  
In the first quadrant, the resistance decreases as the voltage increases and rises again as the voltage returns to zero.  
A similar behavior is observed in the third quadrant.

Under \textbf{pulsed excitation}, the memristor reproducibly switches to about \SI{3.3}{\mega\ohm} at the SET voltage, limited by the compliance current.  
The voltage waveform clips once this limit is reached, confirming current regulation.  
During the subsequent READ phase, the resistance rises to \SI{40}{\mega\ohm}--\SI{100}{\mega\ohm} with noticeable cycle-to-cycle variation, indicating an unstable filament and volatile switching.  
This rapid decay of the conductive state, caused by the low compliance current (\SI{1}{\micro\ampere}), renders the device unsuitable for non-volatile operation under the given conditions.

In \textbf{digital operation}, the observed ON/OFF ratio of approximately 35 is below the benchmark value of 50 reported in \cite{yang2013memristive} for reliable memory applications. This deviation is mainly attributed to the low compliance current, preventing stable filament formation.
Increasing the compliance current would likely improve the switching ratio and device stability but also increase the risk of device failure due to micro-arcing.
The measurement time resolution (\(\Delta t = \SI{0.1}{\second}\)) also limits the detection of the nanosecond-scale switching dynamics (\(t_\mathrm{switch} \approx \SI{1}{\nano\second}\)) reported in \cite{yang2013memristive}.  

For \textbf{analog or neuromorphic operation}, the measured \(R_\mathrm{max}\) exceeding \SI{100}{\mega\ohm} meets the high-resistance benchmark specified in \cite{yang2013memristive}.  
Although the ON/OFF ratio of 35 is below the desired value of 500, the gradual resistance modulation and decay times on the order of seconds are adequate for synaptic and analog computing applications.

The observed instability results from the low compliance current.  
Despite this limitation, the data provide a suitable basis for initial model development.

\section{Structure of Memristor Models} \label{chap: Structure of Memristor Models}

\begin{figure}[b!]
\centering
    \subfloat[]{\includegraphics[width=0.98\linewidth]{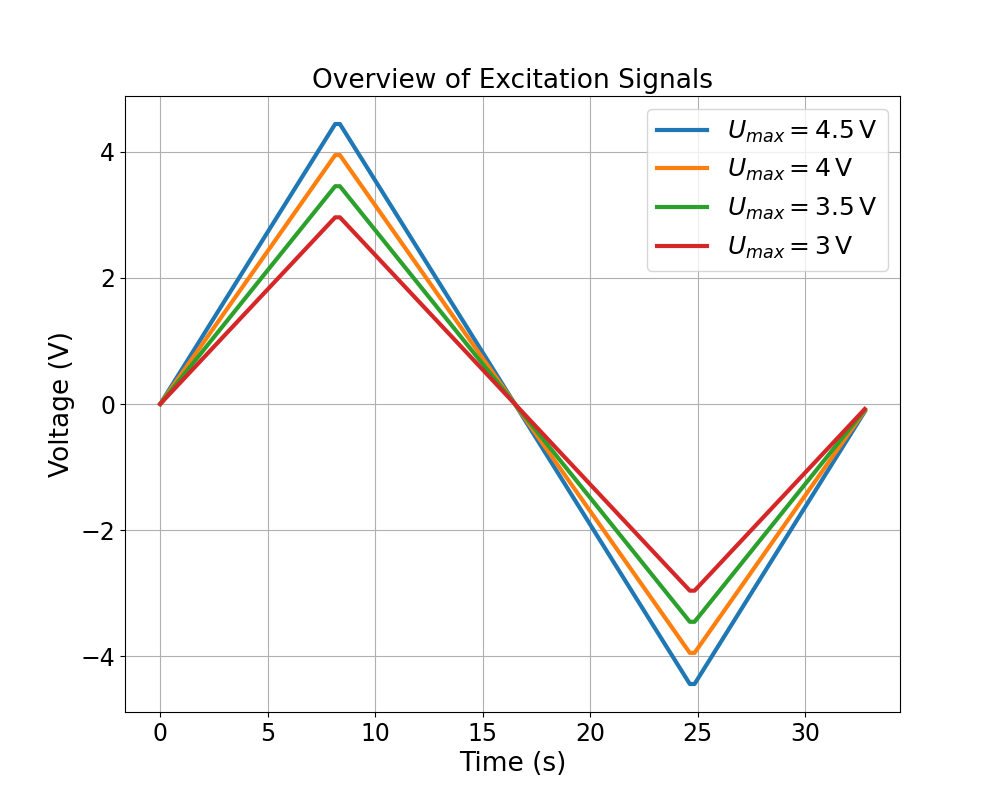}
    \label{pic: yakopcic_excitation}
    }\\

\subfloat[]{\includegraphics[width=0.88\linewidth]{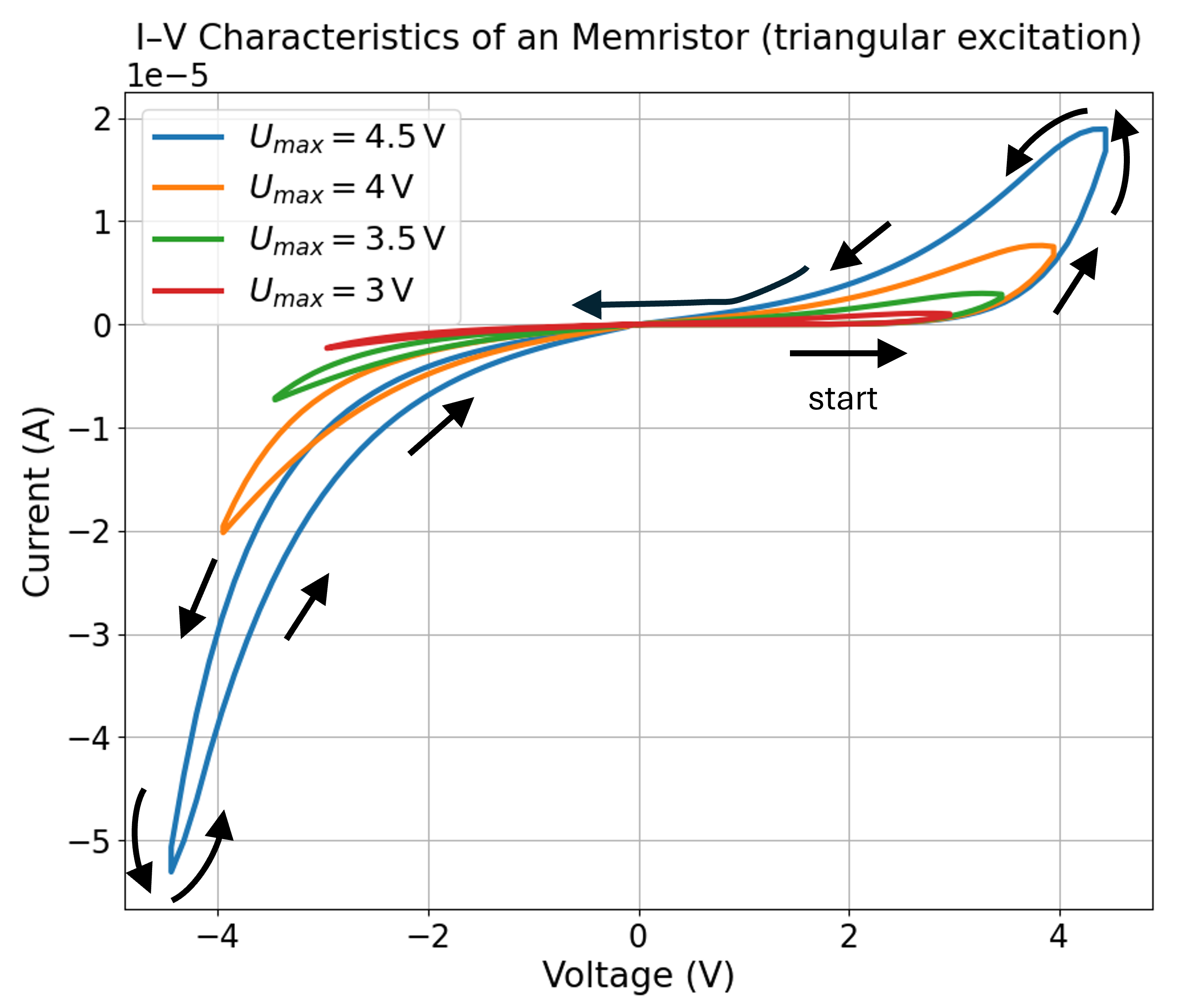}
    \label{pic: yakopcic_iv}
    }
\caption{Simulation results of the selected memristor model without leakage term:
(a) input voltage with varying amplitudes, and
(b) the corresponding I–V characteristics.  
The simulations were carried out using the following model parameters:
$a_{1}=2.5\times10^{-6}$, 
$a_{2}=2.6\times10^{-6}$, 
$b=-0.0025$, 
$A_{p}=0.0025$, 
$A_{n}=0.9$, 
$V_{p}=1.5\,\mathrm{V}$, 
$V_{n}=1.0\,\mathrm{V}$, 
$w_{0}=0.01$, 
$p=1$.}
\label{fig:yakopcic_model_results}
\end{figure}

The memristor is a dynamic element. 
Its response depends both on its previous resistance state and on the duration, amplitude, and polarity of the input signal. To describe these characteristics mathematically, we need a model that accurately reproduces the measured experimental data.

The behavior of a memristor is typically described by a set of two coupled equations that capture its history-dependent behavior.  
These equations consist of an expression for the current–voltage relationship and another for the internal state dynamics.

The current–voltage relationship is given by
\begin{equation}
    i(t) = G(x, u)\,u(t),
    \label{eq:GeneralEquation1_VoltageControlled}
\end{equation}
where \(G(x,u)\) denotes the state-dependent conductance, and \(u(t)\) is the applied voltage.  
The function \(G(x,u)\) may exhibit nonlinear dependence on both the internal state \(x\) and the instantaneous input \(u\).

The state dynamic is given by
\begin{equation}
    \frac{dx(t)}{dt} = \eta \, g(x, u)\, f(x, u) + l(x),
    \label{eq:GeneralEquation2_VoltageControlled}
\end{equation}

where \(x(t)\) represents the internal state variable of the memristor.  
Typically, \(x\) is constrained to the interval \(0 \leq x \leq 1\), representing the physical limits of the device’s internal state.

Equation~\eqref{eq:GeneralEquation2_VoltageControlled} describes the temporal evolution of \(x\).  
The right-hand side consists of several terms, each with a distinct physical interpretation:

\begin{itemize}
    \item \textbf{\(\eta\)} is a \textit{direction factor} that specifies whether a positive current increases or decreases the internal state.
    \item \textbf{\(g(x,u)\)} is a \textit{threshold function} that determines when a change in state occurs.  
    Below the threshold, the internal state remains constant.
    \item \textbf{\(f(x,u)\)} is a \textit{window function} that reduces the rate of change of \(x\) near its boundary values, ensuring that \(x\) remains within the defined limits.
        \item \textbf{\(l(x)\)} is a \textit{leakage term} describing relaxation of the internal state, independent of the applied voltage
\end{itemize}

Together, these two equations provide a general mathematical framework to describe the memristive behavior.

\section{Selected Memristor Model} \label{chap: Selected Memristor Model}

The memristor model used in this work is based on the Yakopcic model~\cite{yakopcic2013generalized}.  
According to Yakopcic, this approach is particularly promising because it can be fitted to a wide variety of memristors.

The model introduces a nonlinear current–voltage (I–V) relationship described by a hyperbolic sine function, together with a threshold function and a window function to capture the internal state dynamics of the device. The use of the hyperbolic sine term reflects the nonlinear conduction mechanism across the metal–insulator–metal interfaces.

In this implementation, the original Yakopcic model is slightly modified by replacing its window function with the one proposed by Biolek \cite{biolek2009spice}, which provides similar boundary behavior but requires fewer parameters to identify. Furthermore, the model is extended by introducing a leakage term based on the approach described in~\cite{chen2013synapse}, which accounts for the gradual loss of resistance (state degradation) when no external voltage is applied.

\begin{figure}[t!]
\centering
    \subfloat[]{\includegraphics[width=0.9\linewidth]{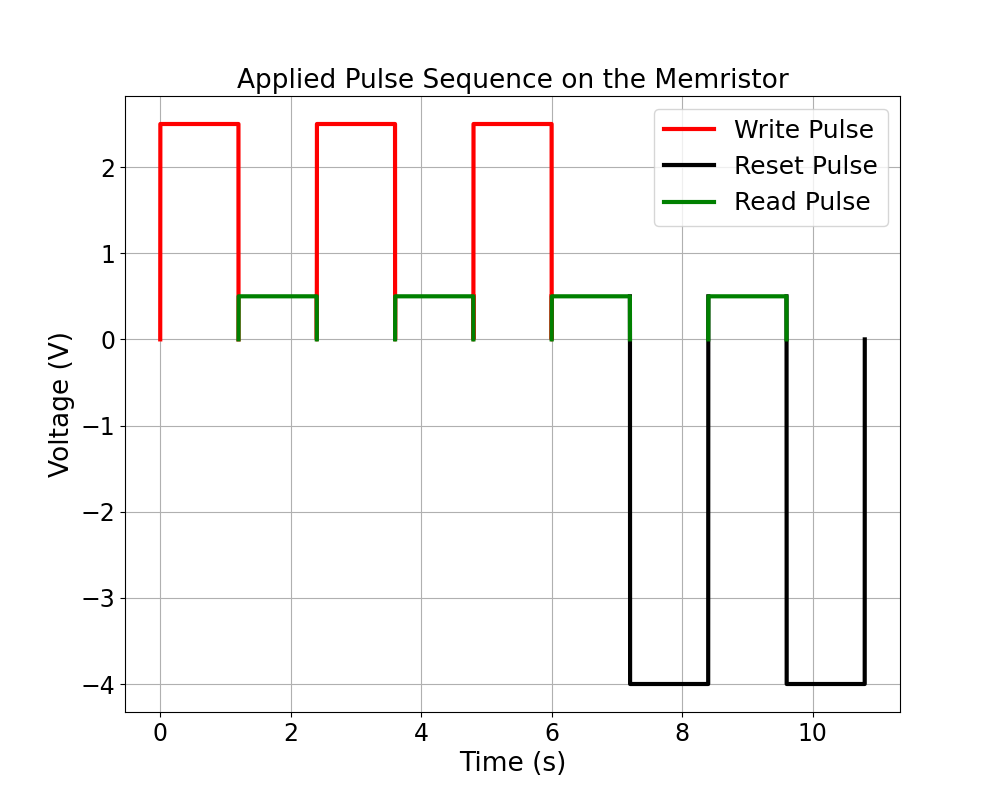}
    \label{pic: yakopcic_excitation_pulses}
    }\\

\subfloat[]{\includegraphics[width=0.9\linewidth]{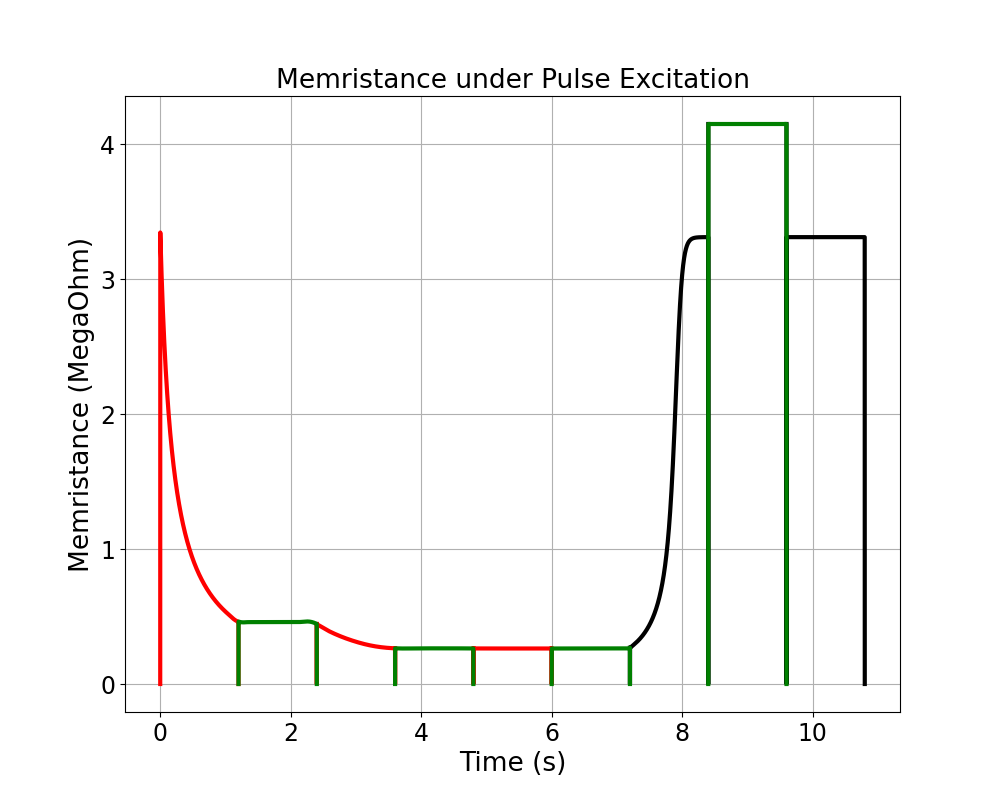}
    \label{pic: yakopcic_iv_pulses}
    }
\caption{Simulation results of the Yakopcic model under pulse-based excitation: 
(a) applied pulse sequence and (b) resulting I–V and state response.}
\label{fig:yakopcic_model_results_pulse}
\end{figure}

The model is presented in detail in the following subsections, and the results are shown in Figure \ref{fig:yakopcic_model_results} for triangular excitation with different amplitudes and in Figure \ref{fig:yakopcic_model_results_pulse} for a selected pulse sequence. The simulations shown do not include the leakage term; therefore, no state changes occur during voltage-free intervals.

\subsection{I–V Relationship}

\begin{figure}[t!]
\centering
    \subfloat[]{\includegraphics[width=0.9\linewidth]{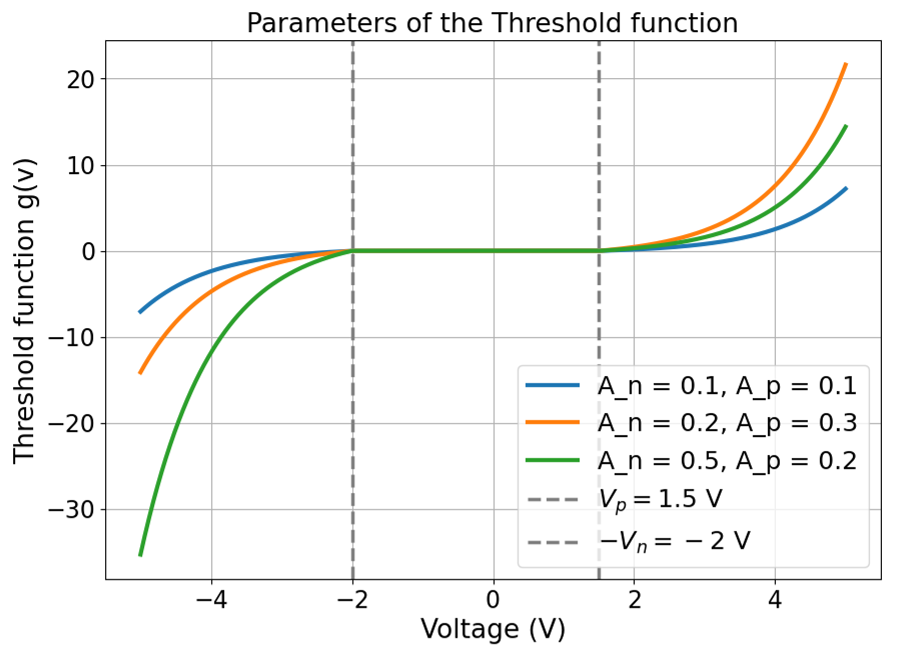}
    \label{pic: Threshold_function}
    }\\

\subfloat[]{\includegraphics[width=0.9\linewidth]{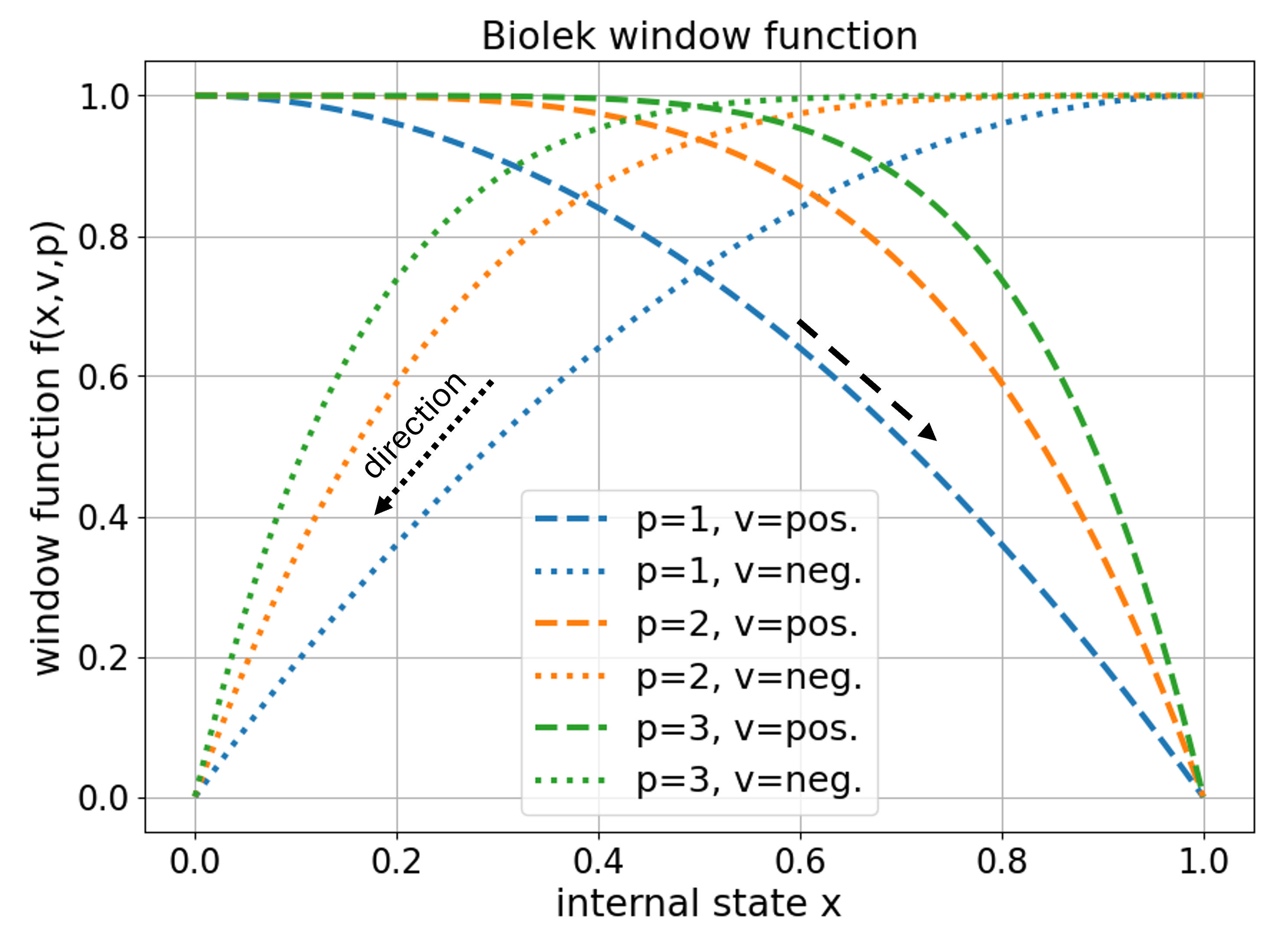}
    \label{pic: bioleks_window_function}
    }
\caption{Functions defining the state dynamics of the proposed model:
(a) Threshold function for different parameter values (b) Biolek window function for various \(p\).}
\label{fig: Threshold_und_window_function}
\end{figure}

The nonlinear current–voltage characteristic is defined as
\begin{equation}
    i(t) =
        \begin{cases}
            a_{1}\,x\,\sinh\!\bigl(b\,u(t)\bigr), & u(t) \ge 0,\\[6pt]
            a_{2}\,x\,\sinh\!\bigl(b\,u(t)\bigr), & u(t) < 0,
        \end{cases}
    \label{eq:sinh_model_eq_1}
\end{equation}
where \(a_i, b > 0\) are constants and \(x\) is the internal state variable. The hyperbolic sine term represents the nonlinear charge transport across the metal–insulator–metal interface. Asymmetry in the I–V characteristics can be modeled through different coefficients \(a_1\) and \(a_2\), allowing the current response to differ for positive and negative voltage polarities.

\subsection{Threshold Function}

The threshold function (Figure~\ref{pic: Threshold_function}) 
\begin{equation}
    g\!\left(u\right) =
    \begin{cases}
        A_p \left( e^{u(t)} - e^{V_p} \right), & u(t) > V_p, \\[6pt]
        - A_n \left( e^{-u(t)} - e^{V_n} \right), & u(t) < -V_n, \\[6pt]
        0, & -V_n \le u(t) \le V_p.
    \end{cases}
\end{equation}
defines the voltage thresholds for state change. 
  
Below the threshold voltages \(V_p\) and \(-V_n\), the internal state remains unchanged, corresponding to \(g(u(t)) = 0\).
Once the applied voltage exceeds these limits, the exponential term becomes active and the state begins to change, with the parameters \(A_p\) and \(A_n\) controlling how rapidly this transition takes place. \newline
Mathematically, the exponential term can be viewed as derived from a difference of hyperbolic sine functions, e.g. \(\sinh(u) - \sinh(V_p)\), which effectively suppresses state changes below the threshold voltage and activates them only beyond it.  

\subsection{Window Function}

To prevent the state variable \(x\) from exceeding its physical limits the Biolek window function (Figure \ref{pic: bioleks_window_function})

\begin{equation}
    f(x,u,p) = 
        \begin{cases}
            1 - x^{2p}, & u > 0, \\[6pt]
            1 - \bigl(x - 1\bigr)^{2p}, & u \leq 0,
        \end{cases}
    \qquad p \in \mathbb{N}.
\end{equation}
is employed. This window function smoothly reduces the rate of change of \(x\) as it approaches its boundaries, ensuring stable and physically meaningful behavior.

\subsection{Leakage Term}

\begin{figure}[t!]
\centering
    \includegraphics[width=0.9\linewidth]{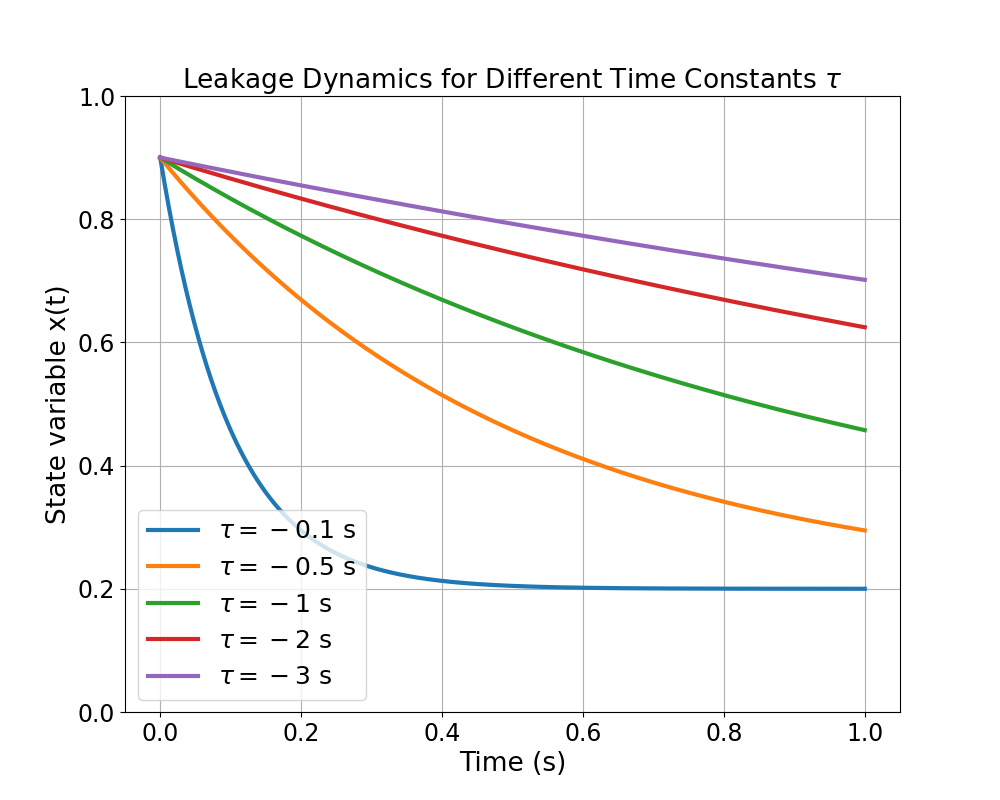}
\caption{Simulated leakage dynamics of the state variable \(x(t)\) for different time constants \(\tau\).  
Parameters: \(x_0 = 0.9\) and \(x_\mathrm{eq} = 0.2\).}
\label{fig: yakopcic_leakage_taus}
\end{figure}

The leakage term
\begin{equation}
    l(x) = -\frac{x - x_{\text{eq}}}{\tau}
\end{equation}
introduces a slow relaxation of the internal state toward an equilibrium value \(x_{\text{eq}}\) in the absence of an external excitation, where \(\tau > 0\) is the time constant that determines how fast the internal state decays toward equilibrium.  
The parameter \(x_{\text{eq}}\) represents the equilibrium state and is constrained to the physically meaningful range \(0 \leq x_{\text{eq}} \leq 1\).

The analytical solution for the case of zero input voltage is
\begin{equation}
    x(t) = x_{\text{eq}} + \bigl(x_0 - x_{\text{eq}}\bigr)e^{-t/\tau},
\label{eq: state_change_leakage_term}
\end{equation}
where \(x_0\) is the initial state at \(t=0\).  
This exponential relaxation toward \(x_{\text{eq}}\) is illustrated in Figure~\ref{fig: yakopcic_leakage_taus}, which shows the state evolution for different time constants~\(\tau\).  
Such a leakage mechanism becomes relevant for pulse-based excitation, where the applied voltage periodically returns to zero and the leakage effect influences the device’s retention.


\section{Parameter Fitting and Optimization} \label{chap: Parameter Fitting and Optimization}

To ensure that the selected memristor model accurately reproduces the experimental measurements, the model parameters are optimized using measurement data from a TiO\textsubscript{2}-based memristor.  
For this purpose, the model was implemented in Python, allowing efficient simulation, data processing, and parameter fitting.

\subsection{Data Preprocessing}

\begin{figure}[t!]
\centering
    \subfloat[]{\includegraphics[width=0.82\linewidth]{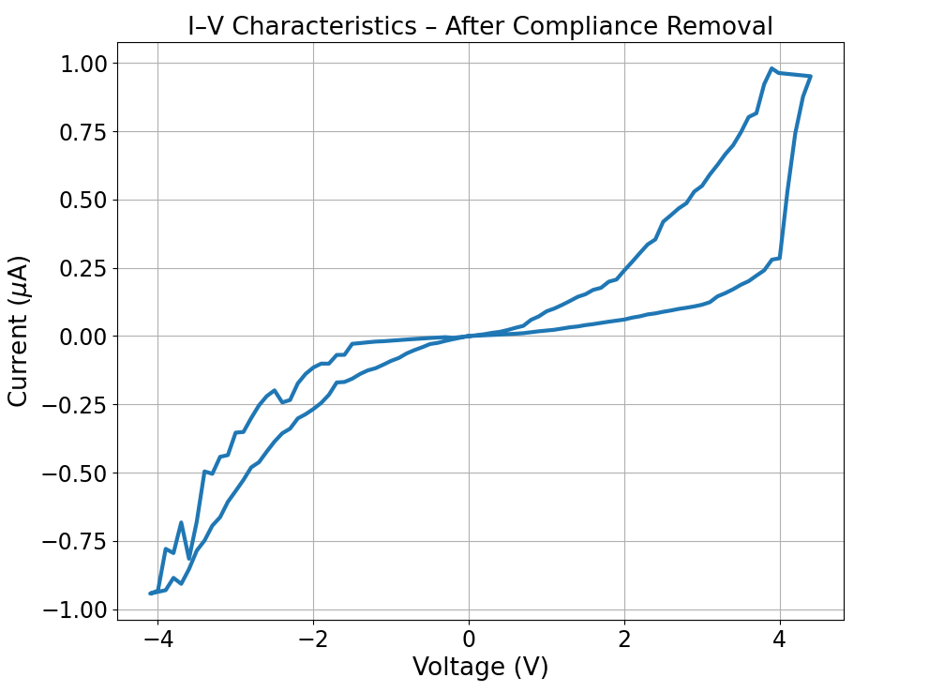}
    \label{pic: I_V_compliance_removal}
    }\\[6pt]
    \subfloat[]{\includegraphics[width=0.82\linewidth]{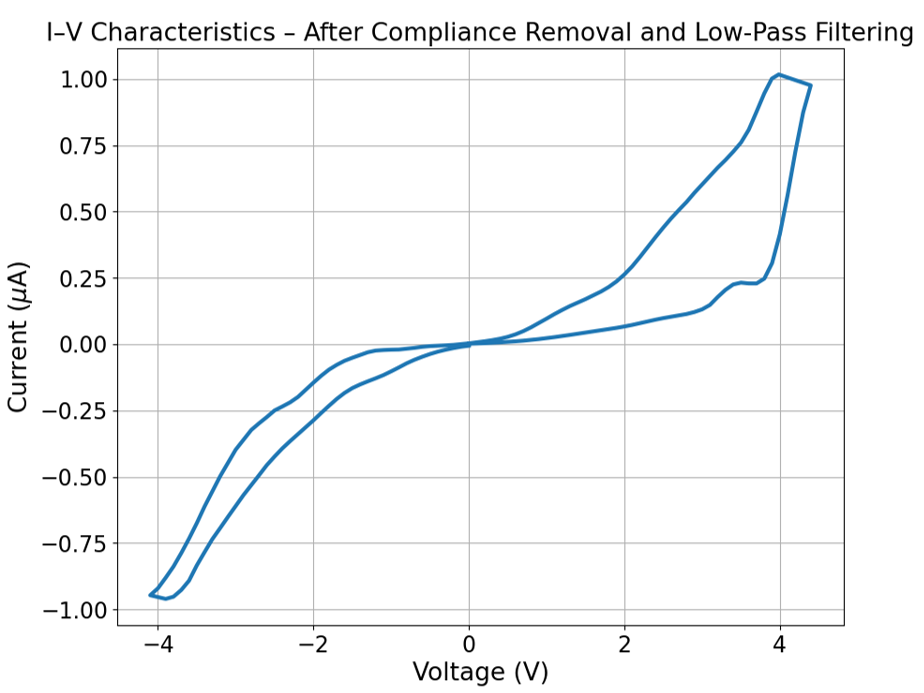}
    \label{pic: I_V_compliance_removal_and_lowpass_filtering}
    }
\caption{Preprocessing of the measured I–V data:  
(a) after compliance current removal,  
(b) after additional low-pass filtering.}
\label{fig: preprocessing_stages}
\end{figure}

Before parameter fitting, the measured data is preprocessed to remove noise and non-modeled effects, as illustrated in Figure~\ref{fig: preprocessing_stages}:
\begin{itemize}
    \item \textbf{Compliance current removal:}  
    The experimental setup limited the maximum current (compliance) to protect the device.  
    Since the memristor model can not handle this effect, all data points affected by the compliance limit were excluded from the fitting dataset.
    \item \textbf{Low-pass filtering:}  
    Applied to the measured current signals to suppress high-frequency noise components in the data.
\end{itemize}

\subsection{Parameter Optimization Strategy}

The parameter optimization aims to minimize the error between the simulated and measured I-V-curves.  
As an objective function, the \textbf{Normalized Mean Squared Error (NMSE)} is used:

\begin{equation}
    \mathrm{NMSE} =
    \frac{\frac{1}{N} \sum\limits_{n=1}^{N} \bigl(\frac{I_{\mathrm{sim}}(n)}{V_{\mathrm{sim}}(n)} 
                -  \frac{I_{\mathrm{meas}}(n)}{V_{\mathrm{meas}}(n)}       \bigr)^{2}}
    {\frac{1}{N} \sum\limits_{n=1}^{N} \frac{I_{\mathrm{meas}}(n)}{V_{\mathrm{meas}}(n)}^{2}},
\end{equation}

where \(I_{\mathrm{meas}}\) is the measured current, \(I_{\mathrm{sim}}\) the simulated current, \(V_{\mathrm{meas}}\) the measured voltage, \(V_{\mathrm{sim}}\) the simulated voltage and \(N\) the total number of samples.


A \textbf{coordinate-descent} strategy is applied to iteratively find the optimal parameter set:

\begin{enumerate}
    \item Select one model parameter and sweep through a range of candidate values.  
    Identify the value that yields the minimum NMSE.
    \item Fix this parameter at its optimal value and move to the next parameter.
    \item Repeat this process for all parameters.  
    After one full cycle, repeat the entire sequence using the newly obtained values.
    \item Stop when the NMSE improvement between two consecutive cycles falls below a predefined tolerance.
\end{enumerate}

This procedure gradually refines all parameters and ensures convergence toward a parameter set that yields the best possible agreement between simulation and measurement.

\subsection{Determination of the Time Constant}

\begin{figure}[b!]
\centering
    \includegraphics[width=0.9\linewidth]{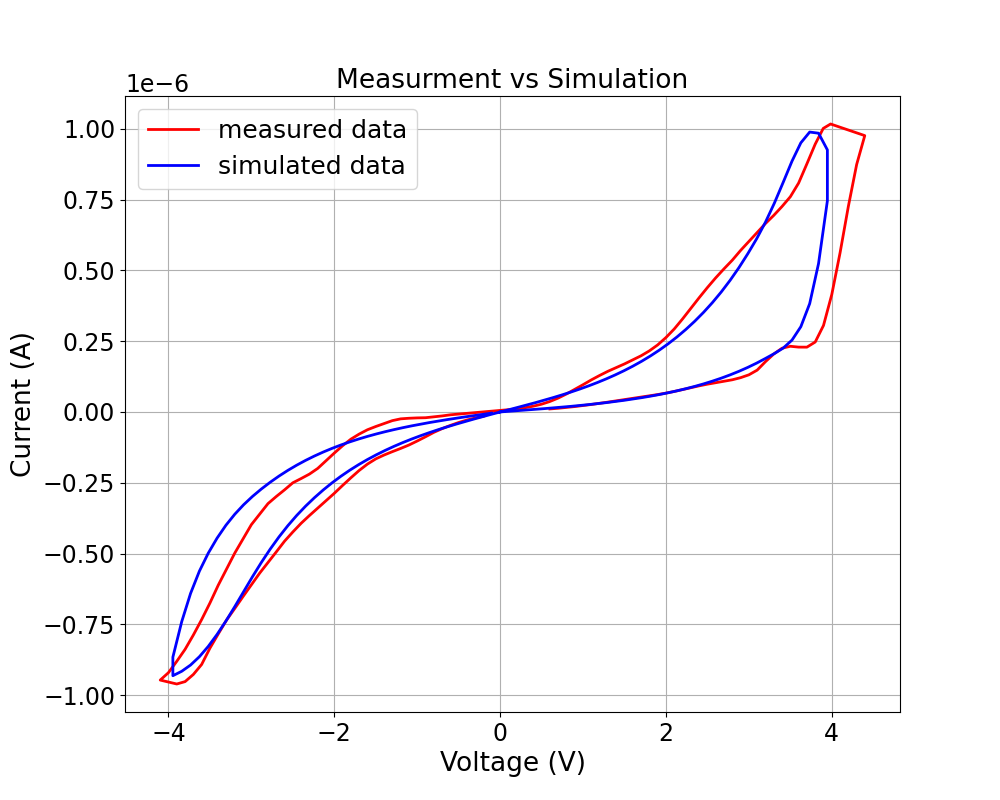}
\caption{Fitted Measurement to the proposed model.}
\label{fig: yakopcic_model_results}
\end{figure}

The time constant~$\tau$ is determined from the leakage dynamics in 
\eqref{eq: state_change_leakage_term} under low or zero applied voltage, where the threshold function~$g(x,u)$ in 
\eqref{eq:GeneralEquation2_VoltageControlled} vanishes.  
In this regime, the I–V relation \eqref{eq:sinh_model_eq_1} becomes approximately linear:
\[
i(t) = a_1 b\,x(t)\,u(t),
\]
and the internal state can be expressed via the measured conductance \(g(t)=i(t)/u(t)\) as
\[
x(t) = \frac{g(t)}{a_1 b}.
\]

Rearranging the analytical leakage equation gives
\begin{equation}
    \tau = -\frac{t_1 - t_0}{\ln\!\left(\frac{x(t_1) - x_{\text{eq}}}{x(t_0) - x_{\text{eq}}}\right)},
\end{equation}
or, equivalently in terms of conductance,
\begin{equation}
    \tau = -\frac{t_1 - t_0}{\ln\!\left(\frac{g(t_1) - g(t_\infty)}{g(t_0) - g(t_\infty)}\right)}.
\end{equation}
If the device reliably returns to its initial state after each cycle (\(x_{\text{eq}}=0\)), this simplifies to
\begin{equation}
    \tau = -\frac{t_1 - t_0}{\ln\!\left(\frac{g(t_1)}{g(t_0)}\right)}.
\end{equation}

For the experimental data (see Figure~\ref{pic: measurement_data_pulse_memristance}), 
\(x_{\text{eq}} = 0\) was assumed, and the read interval corresponds to \(t_1 - t_0 = t_\mathrm{read} = \SI{15}{\second}\).  
Using the averaged resistance values \(\bar{M}(t_0) = \SI{3}{\mega\ohm}\) and \(\bar{M}(t_1) = \SI{108}{\mega\ohm}\), 
the mean time constant is obtained as
\[
\bar{\tau} \approx -\SI{4.2}{\second}.
\]
The negative sign results from the logarithmic formulation and indicates an exponential relaxation toward equilibrium.  
The measurement data further suggest the presence of metastable intermediate states, which are not captured by the leakage term.

\subsection{Evaluation of the Extracted Parameters}

Figure~\ref{fig: yakopcic_model_results} shows the fitted measurement data, and the extracted parameters are summarized in Table~\ref{tab:model_parameters}.  
The model provides a good overall fit to the experimental data.  
Unlike most published results, the parameter \(A_n\) is negative, reflecting the observed decrease in memristance under negative bias (Figure~\ref{pic: measurement_data_i_u_t}).  
The window parameter proved difficult to identify reliably due to the limited temporal resolution of the measurements, which also complicates the estimation of \(\tau\).  
Future optimization could use alternative cost functions, such as minimizing the area between measured and simulated I–V curves, to achieve a more accurate global fit.

\begin{table}[hb!]
\centering
\renewcommand{\arraystretch}{1.1}
\resizebox{0.9\linewidth}{!}{%
\begin{tabular}{|c|c|l|}
\hline
\textbf{Parameter} & \textbf{Value} & \textbf{Description} \\ \hline
$a_1$ & \num{0.0000026} & Fitting coefficient for positive bias \\ \hline
$a_2$ & \num{0.0000012} & Fitting coefficient for negative bias \\ \hline
$b$ & \num{0.85} & Nonlinearity factor of $\sinh(bu)$ \\ \hline
$V_p$ & \SI{1.5}{\volt} & Positive threshold voltage \\ \hline
$V_n$ & \SI{1.0}{\volt} & Negative threshold voltage \\ \hline
$A_p$ & \num{0.0055} & Scaling factor for positive threshold \\ \hline
$A_n$ & \num{-0.0012} & Scaling factor for negative threshold \\ \hline
$\eta$ & \num{1} & Polarity factor \\ \hline
$p$ & \num{1} & Exponent of the Biolek window function \\ \hline
$\tau$ & \SI{4.2}{\second} & Leakage time constant \\ \hline
$x_{\text{eq}}$ & \num{0} & Equilibrium state \\ \hline
\end{tabular}
}
\vspace{4pt}
\caption{Parameters of the proposed model.}
\label{tab:model_parameters}
\end{table}

\begin{figure}[b!]
\centering
\begin{minipage}{0.95\linewidth}
\begin{lstlisting}
***********************************************
*       SPICE Model for volatile memristors
***********************************************

* ------------- Parameters  ---------------
.param a_1 = 0.0000026, a_2 = 0.0000012 
.param b = 0.85
.param V_p = 1.5, V_n = 1.0
.param A_p = 0.0055, A_n = -0.0012
.param eta = 1
.param p = 1
.param tau = 4.2, x_eq = 0
.param x_0 = 0

* ------------- Model Equations ---------------
* I-V relation (first Equation)
.func IV_Memristor(V_mem, x) = if(V_mem>=0, a_1*x*sinh(b*V_mem), a_2*x*sinh(b*V_mem))

* Biolek window function
.func window(V_mem, x) = 1 - pow((x - u(-V_mem)), 2*p)

* Threshold-Function
.func G_Threshold(V_mem) = if(V_mem <= V_p, (if(V_mem >= -V_n, 0, -A_n*(exp(-V_mem) - exp(V_n)))), (A_p*(exp(V_mem) - exp(V_p))))

* Leakage Term
.func Leakage(x) = -(x+x_eq)/tau

* State equation (second Equation)
.func dxdt(V_mem, x) = eta*window(V_mem,x)*G_Threshold(V_mem) + Leakage(x)

* ------- Subcircuit Construction -------
.SUBCKT MEM_YAKOPCIC_LEAKAGE TE BE

* Initial value of the internal state
.IC V(x_intern) = x_0 

* First behavioral current source
G_Memristor TE BE value = {IV_Memristor( V(TE, BE), V(x_intern, 0))}

* Second behavioral current source
G_differential_state_Memristor 0 x_intern value = {dxdt(V(TE, BE), V(x_intern, 0))}

* Integration capacitor
Cx x_intern 0 {1}

.ENDS MEM_YAKOPCIC_LEAKAGE
\end{lstlisting}
\end{minipage}
\caption{Translating the proposed memristor model into SPICE Code.}
\label{fig: Spice_Code_Yoglekar}
\end{figure}

\section{SPICE Implementation} \label{chap: SPICE Modelling}
\begin{figure}[h!]
\centering
    \includegraphics[width=0.98\linewidth]{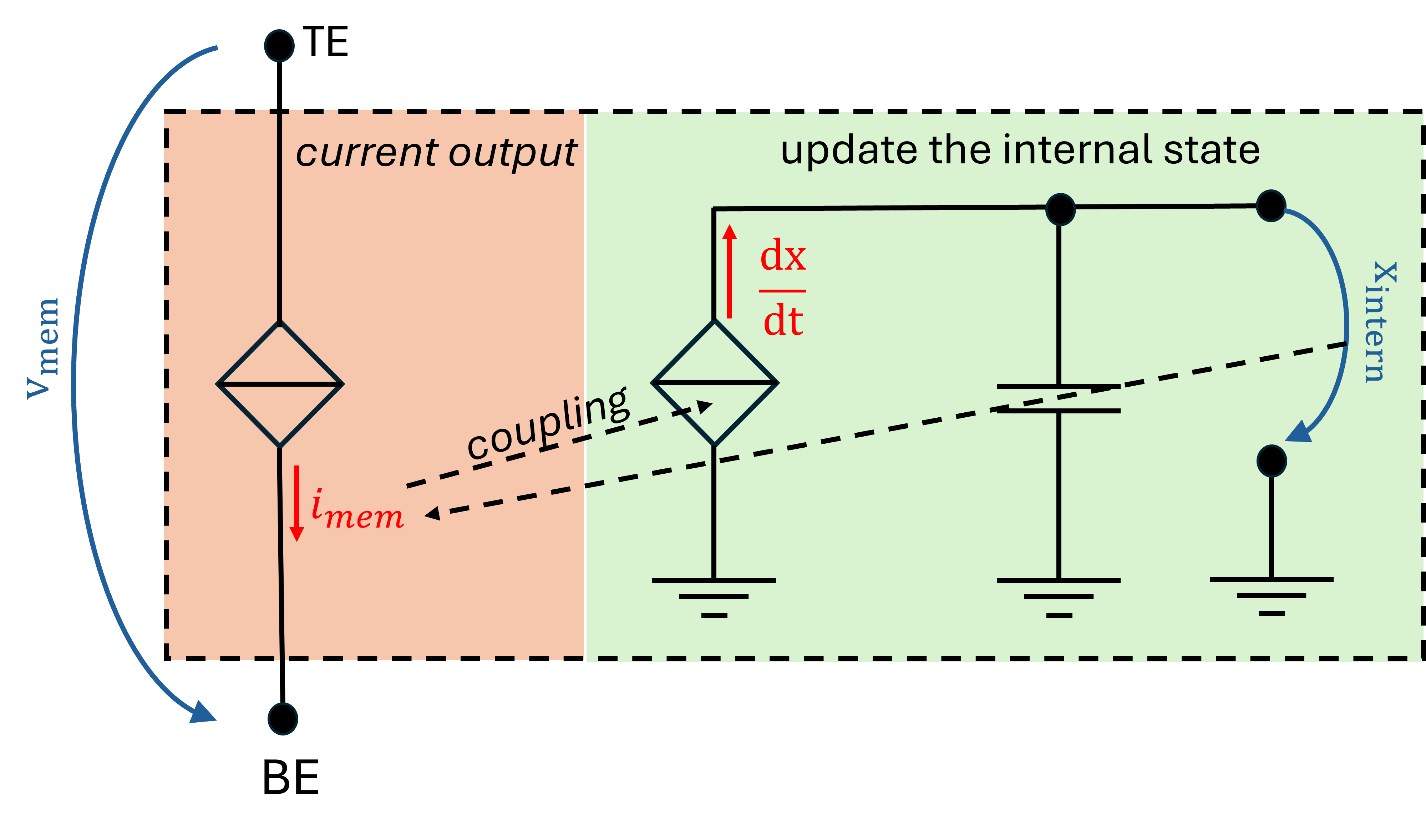}
\caption{SPICE subcircuit of the memristor model.}
\label{fig: SPICE_subcircuit}
\end{figure}

This section describes the translation of the developed memristor model into a SPICE-compatible subcircuit. For that the previously developed model equations and identified device parameters are translated into SPICE code (Figure~\ref{fig: Spice_Code_Yoglekar}), resulting in a reusable SPICE subcircuit (Figure~\ref{fig: SPICE_subcircuit}) that reproduces the dynamic behavior of the memristor based on the extracted parameters.

The subcircuit (Figure~\ref{fig: SPICE_subcircuit}) consists of two behavioral current sources.  
The left-hand source models the current–voltage relation \eqref{eq:GeneralEquation1_VoltageControlled} between the terminals TE and BE.  
The right-hand source implements the state equation \eqref{eq:GeneralEquation2_VoltageControlled}, generating the time derivative \(\dot{x}(t)\); the role of this source is purely mathematical. 
A capacitor of \SI{1}{\farad} integrates this signal, providing the internal state as a pseudo voltage \(x(t)\).  

The dashed lines in Figure~\ref{fig: SPICE_subcircuit} indicate the bidirectional coupling between both equations:  
the internal state \(x(t)\) influences the I–V behavior, while the memristor current feeds back into \(\dot{x}(t)\).  
This coupling enables the complete dynamic representation of the device within SPICE.

\section{Model Evaluation} \label{chap: Model Evaluation}
This section evaluates the proposed SPICE model at the circuit level. First, the memristor subcircuit is integrated into a SPICE testbench that reproduces the experimental measurement setup, including the behavior of the source measurement unit (SMU). The simulated device characteristics are then compared with the measured data to validate the accuracy of the proposed model. Finally, the applicability of the model is demonstrated by integrating it into a leaky integrate-and-fire neuron circuit, illustrating its suitability for neuromorphic circuit simulations.

\begin{figure}[b!]
\centering
    \includegraphics[width=0.93\linewidth]{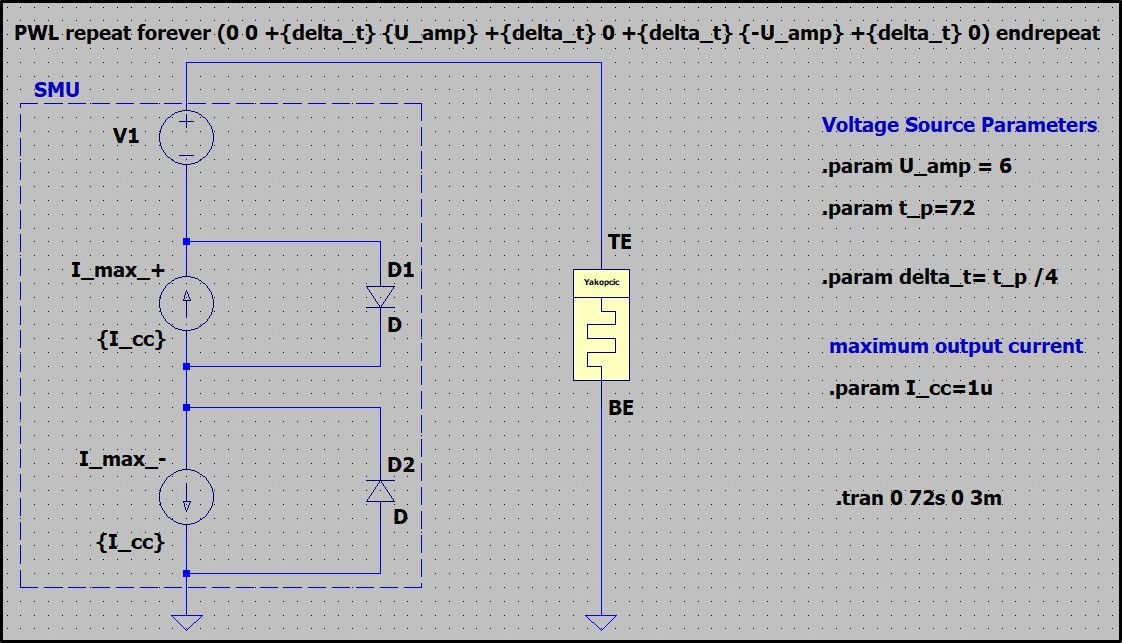}
\caption{SPICE Testbench.}
\label{fig: SPICE_Testbench}
\end{figure}

\subsection{SPICE Testbench}
In order to evaluate the proposed model, a SPICE testbench was created in which the memristor can be excited by either voltage pulses or triangular waveforms.
This setup mimics the measurement system, where a defined compliance current limits the maximum current through the device (Figure~\ref{fig: SPICE_Testbench}). 
This limitation is imposed by the external circuit, as the model itself does not include a current-limiting mechanism, ensuring consistency with the measurement data that are valid only within the same current range.

\subsection{SPICE Model Validation} \label{chap: Validation}

\begin{figure}[b!]
\centering
    \subfloat[]{\includegraphics[width=0.87\linewidth]{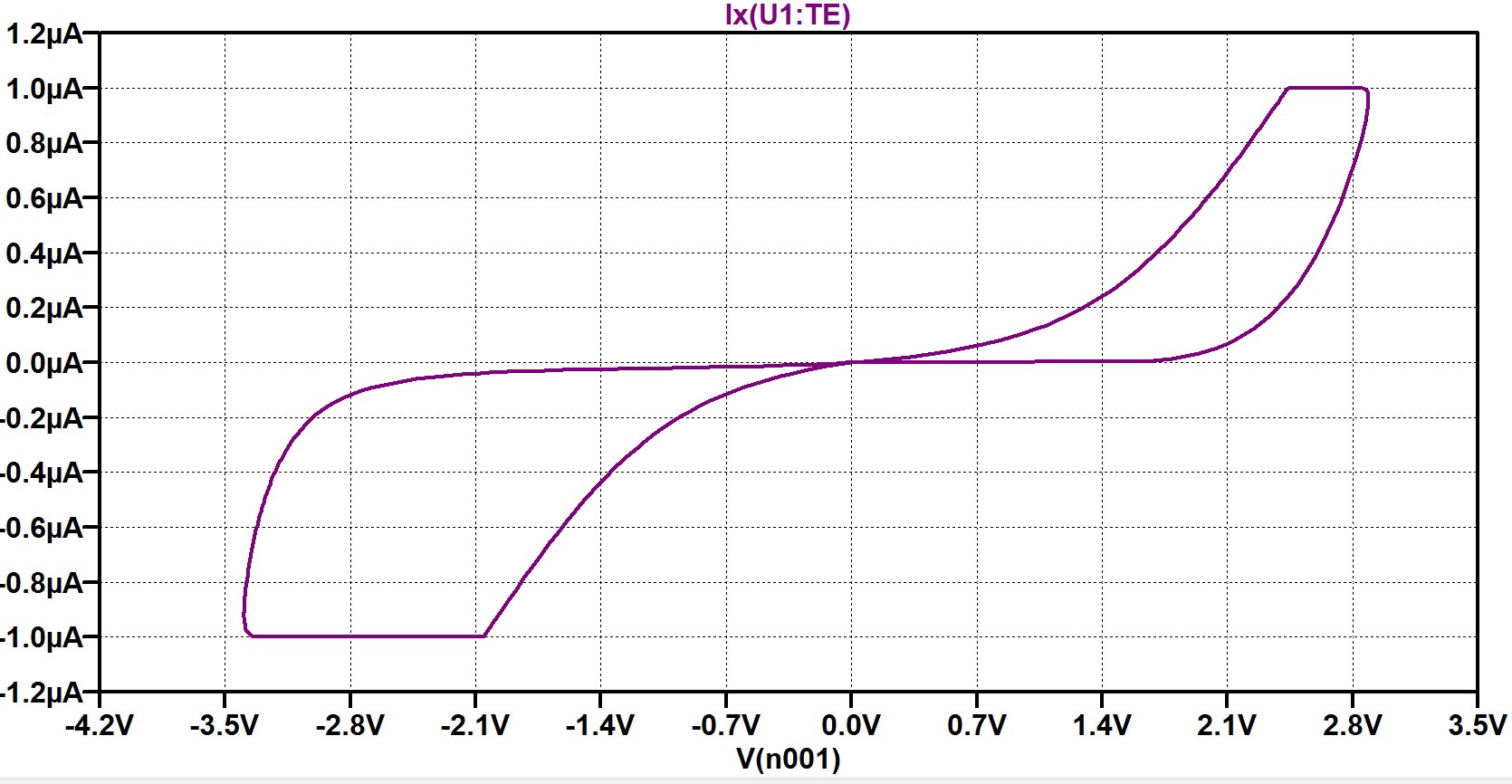}
    \label{pic: SPICE_I_U_Kennlinie}
    }\\[6pt]
    \subfloat[]{\includegraphics[width=0.87\linewidth]{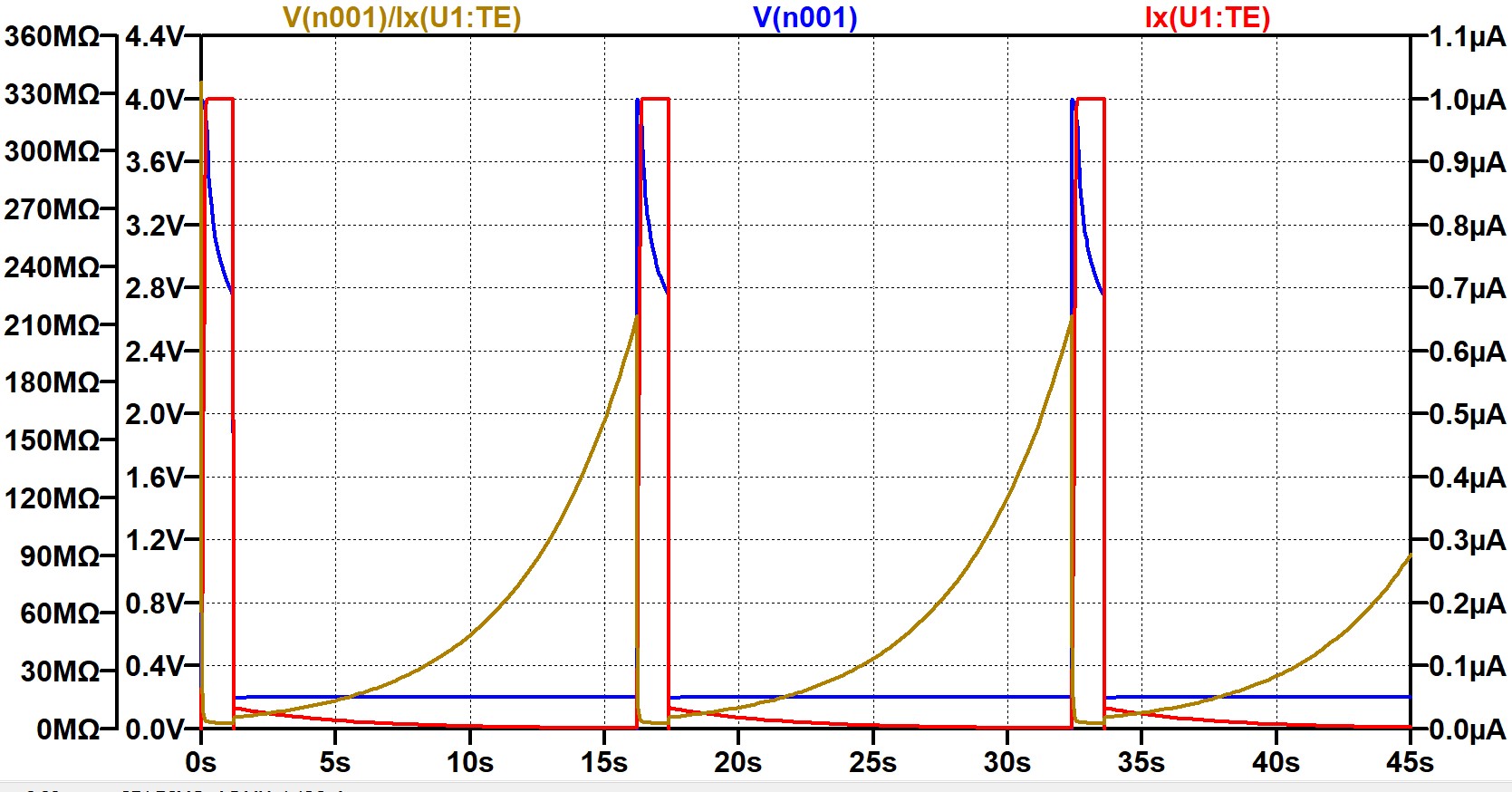}
    \label{pic: SPICE_Pulses}
    }
\caption{SPICE-Simulation results of the proposed memristor model:  
(a) simulated I–V characteristics,  
(b) simulated response under pulsed excitation.}
\label{fig: Results of the SPICE-Testbench}
\end{figure}
Using the optimized parameter set from Table~\ref{tab:model_parameters}, the SPICE model was simulated under the same excitation conditions as the measurements. Figure~\ref{fig: Results of the SPICE-Testbench} shows the simulated responses for triangular and pulsed excitation, which are compared with the corresponding measurement data in Fig.~\ref{pic: measurement_data_I_V} and Fig.~\ref{pic: measurement_data_pulse_memristance}.

For the \textbf{triangular excitation}, the simulated I--V characteristic reproduces the measured hysteresis loop well below the compliance-current limit. Deviations occur mainly in the compliance region, where the simulation exhibits a horizontal segment due to the idealized current-limiting mechanism. In contrast, the measured data show voltage fluctuations (Fig.~\ref{pic: measurement_data_i_u_t}), indicating ongoing resistance variations that are not captured by the deterministic model. These deviations are therefore attributed to stochastic effects in the real device.

For \textbf{pulsed excitation}, the simulation captures the volatile switching behavior of the device, including the resistance increase during the write pulse and the subsequent relaxation during the read pulse. Due to the nonlinear I--V relationship in Eq.~\eqref{eq:sinh_model_eq_1}, different pulse amplitudes result in different apparent resistance values. This behavior is also observed in the measurement data in Fig.~\ref{pic: measurement_data_pulse_memristance}, where a resistance jump occurs between the set and read pulses before the resistance gradually relaxes. In contrast to the simulation, the measured decay exhibits noticeable fluctuations, indicating stochastic device behavior that is not captured by the deterministic model.

Overall, the SPICE implementation captures the main qualitative features of the measured device behavior, including the nonlinear I--V characteristic, the current-limited switching behavior, and the volatile resistance relaxation. The remaining deviations are mainly caused by stochastic fluctuations of the real device, which are not included in the deterministic model.

\subsection{Neuromorphic Application Example}

\begin{figure}[b!]
\centering
    \subfloat[]{\includegraphics[width=0.93\linewidth]{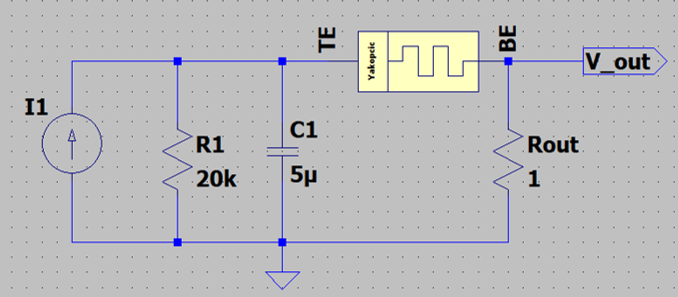}
    \label{pic: LIF_circuit}
    }\\[6pt]
    \subfloat[]{\includegraphics[width=0.93\linewidth]{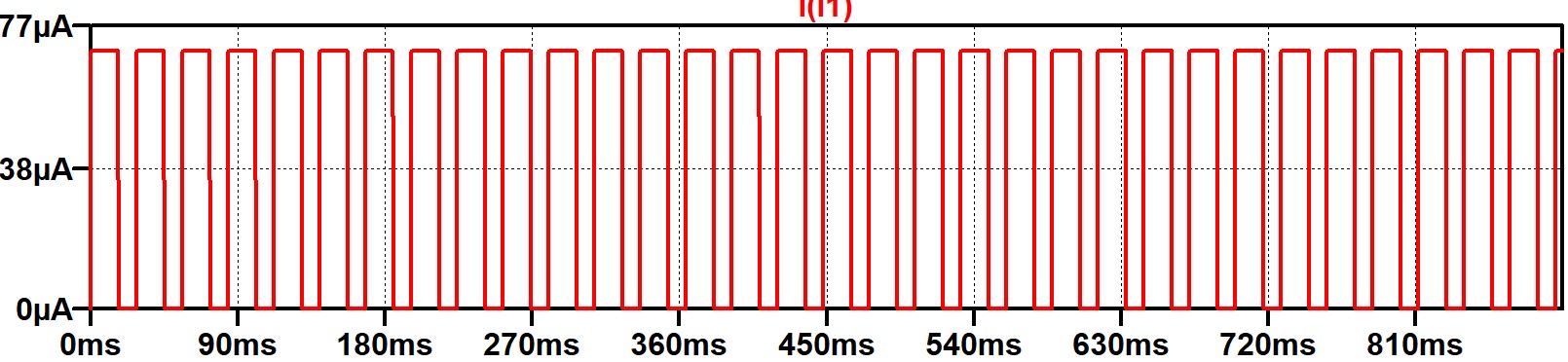}
    \label{pic: LIF_input}
    }\\[6pt]
    \subfloat[]{\includegraphics[width=0.93\linewidth]{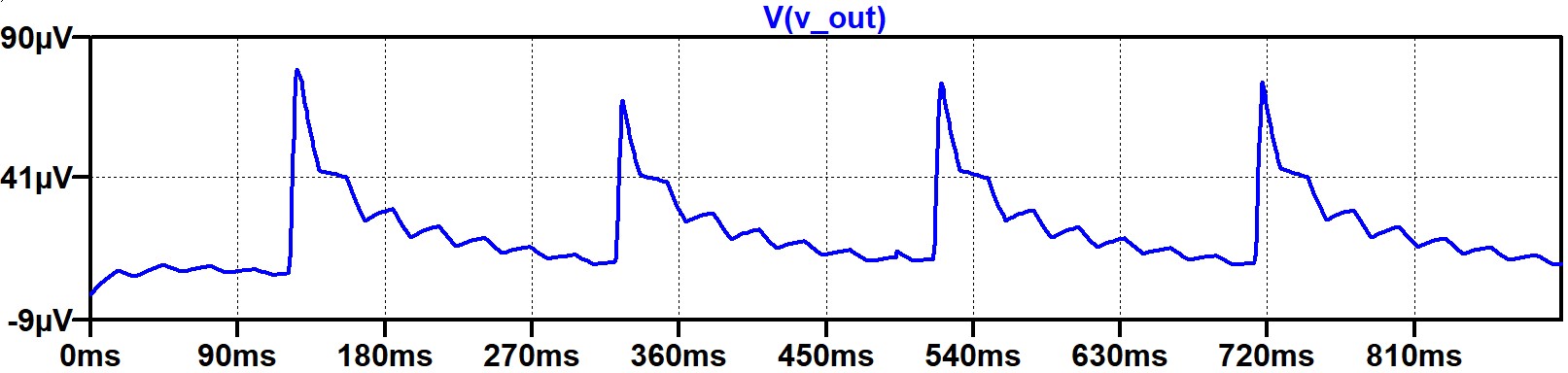}
    \label{pic: LIF_output}
    }
\caption{Example application of the proposed memristor model in a leaky integrate-and-fire neuron based on \cite{dutta2017leaky}:
(a) Modified SPICE circuit in which the original MOSFET is replaced by the proposed memristor model,
(b) applied input current pulses,
(c) simulated output voltage showing spike generation.}
\label{fig: LIF_example}
\end{figure}

To demonstrate the applicability of the proposed model, it was integrated into the leaky integrate-and-fire neuron presented in \cite{dutta2017leaky}. The original circuit consists of a current source, resistor, capacitor, and MOSFET connected in parallel. In the proposed implementation, the MOSFET is replaced by the developed memristor model, which serves as the threshold element (Figure~\ref{pic: LIF_circuit}). Since the memristor produces a current spike during switching, an additional resistor is used to convert this current into the output voltage.

The circuit was excited by a train of current pulses (Figure~\ref{pic: LIF_input}) with an amplitude of \SI{70}{\micro\ampere}, a period of \SI{28}{\milli\second} and an on-time of \SI{17}{\milli\second}. As shown in Figure~\ref{pic: LIF_output}, repeated output spikes with amplitudes of up to approximately \SI{80}{\micro\volt} are generated. The RC network acts as a leaky integrator until the memristor threshold is reached, causing a rapid decrease in resistance resulting in a current spike. Afterwards, the implemented leakage term gradually restores the memristor to its initial high-resistance state, enabling repeated spiking under continuous excitation. Without the leakage term, the memristor would remain in its low-resistance state after switching, preventing repetitive spike generation.

\section{Conclusion and Future Work}

In this work, a volatile TiO\textsubscript{2} memristor was experimentally characterized, modeled, and implemented as a SPICE subcircuit. The Yakopcic memristor model was extended by a leakage term to reproduce the experimentally observed relaxation of the device state toward its initial condition. Comparison with experimental measurements demonstrated that the proposed model accurately reproduces the electrical behavior of the investigated device. Furthermore, the applicability of the model was demonstrated by its integration into a leaky integrate-and-fire neuron circuit.

The presented workflow provides a systematic approach for deriving a SPICE model of volatile memristors directly from experimental characterization data.

Since the relaxation process exhibits stochastic behavior, the equilibrium state \(x_{\mathrm{eq}}\) should be considered a stochastic variable in future model refinements. Experimentally, increasing the compliance current could further stabilize the conductive filament and improve retention. Future work will therefore focus on incorporating stochastic effects to account for device variability, including differences between individual memristors and cycle-to-cycle fluctuations, as reported in \cite{xie2024data}. These refinements will require measurements on a larger number of devices.

Consequently, the neuron dynamics may be analyzed using the theory of Van der Pol oscillators \cite[Chap.~7]{thompson2002nonlinear}, providing a useful analytical framework for future investigations of memristor-based neuromorphic circuits and larger networks.

\section*{Acknowledgment}
This work was funded by the Carl Zeiss Foundation via the project MemWerk (Contract No.~P2018-01-002). \newline
The authors would like to thank Jonas Schneegaß and Joachim Döll from the Center of Micro- and Nanotechnology (ZMN), a DFG-funded core facility at TU Ilmenau, for their technical support and help with some of the experiments.

\printbibliography[title={References}]

\end{document}